\documentclass[12pt]{article}
\usepackage[normalem]{ulem}
\usepackage{amsfonts,slashed}
\usepackage{arydshln}
\usepackage[greek,english]{babel}
\usepackage{cancel}
\usepackage{upgreek}
\usepackage{float}
\usepackage{url}
\usepackage{latexsym}
\usepackage{mathtools}
\usepackage{amsfonts}
\usepackage{amsmath, amscd}
\usepackage{amssymb}
\usepackage{epsfig}
\usepackage{latexsym,amssymb}
\usepackage{amsmath,amssymb,amsthm}
\usepackage{hyperref}
\usepackage{fontenc}
\usepackage{graphicx,tikz,shortvrb}
\usepackage{multirow}
\usepackage{xfrac}
\usepackage[T1]{fontenc}
\usepackage{textcomp}
\usepackage{lmodern}
\usepackage{stmaryrd}
\usepackage{cite}
\usepackage{mathrsfs,mathabx}
\usepackage[margin=20pt,small]{caption}
\usepackage{subcaption}
\usepackage{bbm}
\usepackage{accents}
\usepackage{geometry}
\usepackage{newlfont}
\usepackage{tabularx}

\usepackage{ifpdf}
\ifx\pdfoutput\undefined
   \pdffalse
   \usepackage{cite}
 \else
   \pdfoutput=1
   \pdftrue
\fi
\DeclareFontFamily{OMS}{rsfs}{\skewchar\font'60}
\DeclareFontShape{OMS}{rsfs}{m}{n}{<-5>rsfs5 <5-7>rsfs7 <7->rsfs10 }{}
\DeclareSymbolFont{rsfs}{OMS}{rsfs}{m}{n}
\DeclareSymbolFontAlphabet{\Scr}{rsfs}

\numberwithin{equation}{section}
\def\be{\begin{equation}}
\def\ee{\end{equation}}
\def\ba{\begin{array}}
\def\ea{\end{array}}

\newcommand{\bea}{\begin{eqnarray}}
\newcommand{\eea}{\end{eqnarray}}

\newlength{\dhatheight}

\begin{document}
\begin{titlepage}

\begin{center}
	{\LARGE \bf
	Timelike Supersymmetric Solutions of\\ D=3, N=4 Supergravity \\[1cm]}

	{ \bf 
	Nihat Sadik Deger\,$^{a,b,c,}{\!}$
		\footnote{\tt sadik.deger@bogazici.edu.tr},
	Ceren Ayse Deral\,$^{a,}{\!}$
		\footnote{\tt ceren.deral@bogazici.edu.tr}
	
		 \vskip .8cm}
	
	{\it ${}^a$ Department of Mathematics, Bogazici University, Bebek, 34342, Istanbul, T\"urkiye}\\[1.5ex] \ 
	{\it ${}^b$ Feza Gursey Center for Physics and Mathematics, Bogazici University, \\ Kandilli, 34684, Istanbul, T\"urkiye}\\[1.5ex] \ 
 {\it ${}^c$ Erwin Schr\"odinger International Institute for Mathematics and Physics, \\ University of Vienna, Boltzmanngasse 9, 1090, Vienna, Austria} \\[1.5ex] \
	 \\
	
\end{center}
\vfill

\begin{center}
	\textbf{Abstract}
	
\end{center}
\begin{quote}
We study timelike supersymmetric solutions of a $D=3, N=4$ gauged supergravity using Killing spinor bilinears method and prove that AdS$_3$ is the only solution within this class. We then consider the ungauged version of this model. It is found that for this type of solutions, the ungauged theory effectively truncates to a supergravity coupled to a sigma model with a 2-dimensional hyperbolic target space $\mathbb{H}^2$, and all solutions can be expressed in terms of two arbitrary holomorphic functions. The spacetime metric
is a warped product of the time direction with a 2-dimensional space, and the warp factor is given in terms of the K\"ahler potential of $\mathbb{H}^2$. We show that when the holomorphic function that determines the sigma model scalar fields is not constant, the metric on the sigma model target manifold becomes part of the spacetime metric. We then look at some special choices for these holomorphic functions for which the spacetime metric and the Killing spinors are only radial dependent. 
We also derive supersymmetric null solutions of the ungauged model which are pp-waves on the Minkowski spacetime.

\end{quote}
\vfill
\setcounter{footnote}{0}

\end{titlepage}

%


\section{Introduction} \label{sec:Intro}
Importance of supersymmetric solutions in the study of supergravity theories is well established. A prime example is the p-brane solutions (see \cite{Stelle:1998xg} for a review), which led to many important developments in String/M-theory. Finding such solutions is easier compared to trying to solve second order field equations since supersymmetry gives first order constraints to be satisfied. Yet, discovering new examples could still be a challenge unless one looks for solutions with high amount of symmetry. A systematic way of constructing them was developed by Tod \cite{Tod:1983pm, Tod:1995jf}, which is based on bilinear tensors built from a Killing spinor. Using BPS equations, one then derives several algebraic and differential conditions, which restrict the form of the metric as well as the physical fields of the model for such a solution. One of these bilinears is a Killing vector, which has to be either null or timelike and can be used to classify them. Although this is a powerful method and has been applied to many supergravities in different dimensions, it is, in general, not possible to solve all these conditions in full generality. As the dimension gets lower, the problem of explicitly constructing all possible superymmetric solutions becomes relatively easier, since the fact that there is less choice for coordinate dependence of physical fields results in a more manageable set of differential equations as was illustrated for 3-dimensional supergravities in \cite{Gibbons:2008vi, Deger:2010rb, Deger:2013yla, Deger:2016vrn, deBoer:2014iba, Deger:2015tra}.

The model that we will consider in this paper is a particular truncation of a $D=3, N=4$, $SO(4)$ gauged supergravity \cite{Deger:2014ofa} which is connected to $D=6$, $N=(1,0)$ supergravity coupled to a single chiral tensor multiplet  \cite{Nishino:1986dc} by a consistent 3-sphere reduction. 
Since the dimensional reduction is consistent, any solution in $D=3$ can be uplifted to a solution in $D=6$, which gives a strong motivation to look for its supersymmetric solutions. In \cite{Deger:2019jtl}, two examples were found, and the uplift of one of them gave rise to a new dyonic string distribution. A comprehensive study of supersymmetric solutions of this theory was initiated in \cite{Deger:2024xnd} using Tod's Killing bilinears method \cite{Tod:1983pm, Tod:1995jf}, but only the null Killing vector case was worked out explicitly. In addition to those found in \cite{Deger:2019jtl}, several new ones were obtained, some resulting in novel configurations in $D=6$ such as rotating AdS$_3\times$S$^3$ \cite{Deger:2024xnd}. In this paper, we complete the work of \cite{Deger:2024xnd} by studying the timelike Killing vector case.
Rather surprisingly, we find that the gauged model admits only AdS$_3$ as a supersymmetric solution in this class. We then consider the ungauged version of this model. In this case, for timelike solutions, BPS conditions set 2 out of 4 scalars and all vector fields to zero, and the model effectively reduces to a supergravity coupled to a sigma model \cite{Marcus:1983hb, deWit:1992psp, Howe:1995zm}. The remaining scalar fields $(\rho, \theta)$  describe a sigma model with a hyperbolic target space 
$\mathbb{H}^2=SU(1,1)/U(1)$:
\begin{align}
    \mathscr{L}_{\textrm{sigma model}} =     - \frac{1}{2} \sqrt{-g}\, \left[  \partial_{\mu} \rho \, \partial^{\mu} \rho 
+ \sinh^{2}{\rho} \, \partial_{\mu} \theta \, \partial^{\mu} \theta \right]. \label{lagtrunc}
\end{align}
It turns out that all timelike supersymmetric solutions of the ungauged model are characterized
by two arbitrary holomorphic functions, and the only effect of the scalars on the spacetime metric is the appearance of the K\"ahler potential of the $\mathbb{H}^2$ as a warp factor. These features are almost identical with the static supersymmetric solutions obtained in \cite{Howe:1995zm} for the $D=3, N=(2,0)$ ungauged supergravity coupled to a sigma model with an arbitrary K\"ahler target space, except that we have an extra independent holomorphic function that determines the Killing spinors. Moreover, the fact that our sigma model target manifold is 2-dimensional gives rise to an interesting possibility, namely, its metric coinciding with the space part of the solution metric. Indeed, we show that for any non-constant choice of the holomorphic function that determines scalars $(\rho, \theta)$ in terms of the spacetime coordinates, the spacetime metric can be written as
\begin{align}\label{metricgen}
ds^2_{\textrm{spacetime}} =-dt^2+ e^He^{-3K(\rho)}(d\rho^2+\sinh^2\rho\; d\theta^2) \, .
\end{align}
The function $K(\rho)$ in the warp factor is the non-harmonic part of the K\"ahler potential of the sigma model target space $\mathbb{H}^2$. Whereas,  
$H$ is a harmonic function determined by the 2 holomorphic functions. 
We then consider three different choices for the holomorphic functions so that the spacetime metric and the Killing spinor become only radial dependent. We also investigate the singularity structure of the corresponding solutions and find that as the scalar 
field $\rho$ goes to infinity there is a naked ring shaped singularity.

The organization of this paper is as follows: In the next section, we introduce our model. In Section \ref{3}, we summarize the Killing spinor analysis that was done in \cite{Deger:2024xnd} with additional details that are relevant to the case at hand. In section \ref{4}, we give the solution for the Killing spinors for the timelike Killing vector and derive BPS conditions. In sections \ref{5} and \ref{6}, timelike supersymmetric solutions of our gauged and ungauged theories are constructed, respectively. We conclude in section \ref{7}. In the appendix \ref{a1},  we derive null supersymmetric solutions of the ungauged theory, which was not considered in \cite{Deger:2024xnd} and show that they are pp-waves on the Minkowski spacetime.

\section{The Model}\label{model}
The bosonic Lagrangian of the model that we will study in this paper is  \cite{Deger:2019jtl}
\bea \label{Lagrangian}
\mathscr{L}_3 & = & \sqrt{-g} \left(R  - \frac{1}{2}   \left[ 
\partial_{\mu} \xi_{1}\, \partial^{\mu} \xi_{1}
+ \partial_{\mu} \xi_{2}\, \partial^{\mu} \xi_{2}
+ \partial_{\mu} \rho \, \partial^{\mu} \rho 
+ \sinh^{2}{\rho} \, D_{\mu} \theta  D^{\mu} \theta \right] \right.
\nonumber \\
& & \left. \quad \quad \quad
- \frac{1}{4} \, e^{-2 \xi_{1}} \, \mathcal{F}_{\mu \nu}^{1} \, \mathcal{F}^{1 \, \mu \nu}
- \frac{1}{4} \, e^{-2 \xi_{2}} \, \mathcal{F}_{\mu \nu}^{2} \, \mathcal{F}^{2 \, \mu \nu} - V  \right) 
- \frac{k_0}{2} \, \varepsilon^{\mu \nu \rho} \, \mathcal{A}_{\mu}^{1}  \, \mathcal{F}_{\nu \rho}^{2} \,. 
\label{lag}
\eea
This is a truncation of the \(D=3, \, N=4 \,,\, SO(4)\) gauged supergravity found in \cite{Deger:2014ofa} which comes from 
a consistent 3-sphere reduction of the $D=6$, $N=(1,0)$ supergravity coupled to a single chiral tensor multiplet  \cite{Nishino:1986dc}. The model is a specific member of the general class of $D=3$ gauged supergravities constructed earlier \cite{Nicolai:2003bp,deWit:2003ja,deWit:2003fgi}. The truncation above was obtained in \cite{Deger:2019jtl} and preserves 
\( U(1) \times U(1) \subset SO(3) \times SO(3) \simeq SO(4) \)
symmetry. The model has four scalars \( \left( \xi_{1}, \xi_{2}, \rho, \theta \right) \)
and two Abelian vector fields \( \mathcal{A}_{\mu}^{1,2} \) whose field strengths are
\( \mathcal{F}_{\mu \nu}^{1,2} \). All fields are real and the covariant derivative for the scalar field $\theta$ is defined as
\be
D_{\mu} \theta = \partial_{\mu} \theta + 2 \, g_{0} \, \mathcal{A}_{\mu}^{1} \,.
\label{covder}
\ee
The scalar potential does not depend on the field $\theta$ and is given as
\be
V = - 4 \, g_{0}^{2} \, e^{\xi_{1} + \xi_{2}} \, \cosh{\rho} 
+ 2 \, g_{0}^{2} \, e^{2 \xi_{1}} \, \sinh^{2}{\rho}
+ \frac{k_{0}^{2}}{2} \, e^{2 \left( \xi_{1} + \xi_{2} \right)} \,.
\label{pot}
\ee
This potential has only one supersymmetric vacuum, which is AdS$_3$ and is located at \cite{Deger:2019jtl}
\begin{equation}
\rho=0 \, , \qquad e^{\xi_1}= e^{\xi_2}=\frac{2g_0}{k_0} \, .
\label{AdSvacuum}
\end{equation}
The superpotential $W$ of the model is given by
\begin{equation}
\label{superpotential}
W=\frac{e^{\xi_2}}{2}\,\left(-2\,g_0+k_0\,e^{\xi_1}\right)-g_0\,e^{\xi_1}\,\cosh\rho
\, .
\end{equation}

The supersymmetry variation of the gravitino gives the Killing spinor equation as
\begin{align}
      (\partial_\mu+ \frac{1}{4}\,\omega_\mu^{\,\,\,\, bc}\, \gamma_{bc})\,\zeta_a
+ X_\mu \, \epsilon_{a b} \, \zeta^{b} 
 - \frac{1}{2} \, W  \, \gamma_{\mu} \, \zeta_{a} = 0 \, ,
\label{eq4}
\end{align}
where
\begin{align}
 X_\mu =   \frac{1}{4}  \left( 1-\cosh{\rho} \right)  D_{\mu} \theta 
- 2  \epsilon_{\mu}^{\,\,\,\, \sigma \rho} 
\left( \mathcal{F}^{1}_{\rho \sigma} + \mathcal{F}^{2}_{ \rho \sigma} \right) \, .
\label{X}
\end{align}
Setting supersymmetry variations of the matter fermions to zero leads to the BPS conditions \cite{Deger:2019jtl}:
\bea
 \left( \gamma^{\mu} \, \partial_{\mu} \xi_{1,2} \right) \zeta_{a} 
-  \left( \gamma_{\mu} \, \epsilon^{\mu \sigma \rho}
\, \mathcal{F}_{\rho \sigma}^{1,2} \right) \, \epsilon_{a b} \, \zeta^{b}
+ 2\, \frac{\partial W}{\partial \xi_{1,2} } \zeta_{a} & = & 0 \,,
\label{eq1} \\ \left( \gamma^{\mu} \, \partial_{\mu} \rho \right) \zeta_{a}
+ \sinh{\rho} \, \left( \gamma^{\mu} D_{\mu} \theta \right) \, \epsilon_{a b} \, \zeta^{b}
+ 2\, \frac{\partial W}{\partial \rho }
 \, \zeta_{a} & = & 0 \,.
\label{eq3} 
\eea
In these supersymmetry transformations \( \zeta_{a} \)'s  \( \, (a=1,2) \) are defined
in terms of Majorana spinors $\lambda_A$'s \( \, (A=1,2,3,4) \) as
$\zeta_1= \lambda_1 + i\lambda_3$ and $\zeta_2=\lambda_2+ i\lambda_4$\cite{Deger:2019jtl}. Here, 
$\epsilon_{ab}=-\epsilon_{ba} $ such that $\epsilon^{12}=\epsilon_{12}=-1$.
From \eqref{Lagrangian}, Einstein's field equations are found to be
\bea
R_{\mu\nu} & = & \frac{1}{2} \Big( 
(\partial_{\mu} \xi_{1}) (\partial_{\nu} \xi_{1})
+ (\partial_{\mu} \xi_{2}) (\partial_{\nu} \xi_{2})
+ (\partial_{\mu} \rho) (\partial_{\nu} \rho) 
+ \sinh^{2}{\rho} \, (D_{\mu} \theta) (D_{\nu} \theta) \Big) + g_{\mu\nu} \, V
\label{Eineq} \\
& & + \frac{1}{2} \, e^{-2 \xi_{1}} \left( \mathcal{F}_{\mu \rho}^{1} \, \mathcal{F}^{1}{}_{\nu}{}^{\rho}
- \frac{1}{2} \, g_{\mu\nu} \, \mathcal{F}_{\rho\sigma}^{1} \, \mathcal{F}^{1 \, \rho\sigma} \right)
+ \frac{1}{2} \, e^{-2 \xi_{2}} \left( \mathcal{F}_{\mu \rho}^{2} \, \mathcal{F}^{2}{}_{\nu}{}^{\rho}
- \frac{1}{2} \, g_{\mu\nu} \, \mathcal{F}_{\rho\sigma}^{2} \, \mathcal{F}^{2 \, \rho\sigma} \right) \,,
\nonumber
\eea
and the scalar field equations are
\bea
&&
\nabla_{\mu} \nabla^{\mu} \xi_{1} 
+ \frac{e^{-2 \xi_{1}}}{2}   \mathcal{F}_{\mu \nu}^{1}  \mathcal{F}^{1  \mu \nu} 
+ 4 g_0^2  e^{\xi_1 +\xi_2}  \cosh{\rho} - 4 g_0^2  e^{2 \xi_1}  \sinh^{2}{\rho}
- k_0^2  e^{2(\xi_1 +\xi_2)} = 0 , \label{xi1eqn} \\
&&\nabla_{\mu} \nabla^{\mu} \xi_{2} 
+ \frac{1}{2} e^{-2 \xi_{2}}  \mathcal{F}_{\mu \nu}^{2} \, \mathcal{F}^{2 \, \mu \nu} 
+ 4 g_0^2 \, e^{\xi_1 +\xi_2}  \cosh{\rho} - k_0^2 \, e^{2(\xi_1 +\xi_2)}  =  0 \,, \label{xi2eqn}\\
&&\nabla_{\mu} \nabla^{\mu} \rho
- \frac{1}{2} \, \sinh{2 \rho} \, D_{\mu} \theta D^{\mu} \theta
+ 4 g_0^2 \, e^{\xi_1 +\xi_2} \, \sinh{\rho} - 2 g_0^2 \, e^{2 \xi_1} \, \sinh{2 \rho}  =  0 \,, \\
&& \nabla_{\mu} \Big( \sinh^{2}{\rho} \, D^{\mu} \theta \Big)  =  0 \,. \label{thetaeqn} 
\eea
Finally, vector fields should satisfy
\begin{align}
   &  \nabla_{\mu}  \Big( e^{-2 \xi_{1}} \, \mathcal{F}^{1 \, \mu \nu} \Big) -\frac{k_0}{2} \, \epsilon^{\nu \mu \rho} \, \mathcal{F}_{\mu \rho}^{2}
- 2 g_0 \, \sinh^{2}{\rho} \,  D^{\nu} \theta 
  =  0 \,,
\label{A1eqn} \\ 
&\nabla_{\mu}  \Big( e^{-2 \xi_{2}} \, \mathcal{F}^{2 \, \mu \nu} \Big) 
-\frac{k_0}{2} \,  \epsilon^{\nu \mu \rho} \, \mathcal{F}_{\mu \rho}^{1}  =  0 \,.
\label{A2eqn}
\end{align}

The model \eqref{Lagrangian} has two free parameters $g_0$ and $k_0$ whose ratio determines the location of the AdS vacuum \eqref{AdSvacuum}. The ungauged version of this supergravity \cite{Marcus:1983hb, deWit:1992psp, Howe:1995zm} is obtained by setting $g_0=k_0=0$, but unlike the gauged theory, its solutions can not be embedded to $D=6$ using uplift formulas found in \cite{Deger:2014ofa} since they become divergent. In this limit, it is easy to see that two groups of fields $(\xi_{1,2}, A_\mu^{1,2})$ and $(\rho, \theta)$ decouple from each other and setting either one to zero does not create any inconsistency in the field equations \eqref{xi1eqn}-
\eqref{A2eqn} or  supersymmetry variations \eqref{eq1}-\eqref{eq3}. If the first set is decoupled, then the Lagrangian of the remaining fields matches exactly with the Poincar\'e limit of the $N=2$ gauged supergravity coupled to a sigma model with the same target space constructed in \cite{Deger:1999st}. 

\section{Killing Spinor Analysis}
\label{3}
The general analysis of the supersymmetry conditions \eqref{eq4}-\eqref{eq3} was done in \cite{Deger:2024xnd} using Tod's method \cite{Tod:1983pm, Tod:1995jf}. In this section, we summarize its main findings with a few additional results that will be needed for the timelike case. For more details, we refer to \cite{Deger:2024xnd} and also \cite{Deger:2010rb} where this approach was applied to the $D=3, N=8$ gauged supergravity. In this method, one starts with assuming the existence of one set of Killing spinors
$\lambda_{A}$ \( \, (A=1,2,3,4) \) which we take as commuting and main objects are the spinor bilinears constructed out of them: 
\begin{align}
F^{AB}&=\bar{\lambda}^A\lambda^B=-F^{BA} \, , \\
V_\mu^{AB}&=\bar{\lambda}^A\gamma_\mu\lambda^B=V_\mu^{BA} \, .
\end{align}
Note that both bilinears are real with our conventions\footnote{Three-dimensional tangent space indices $i, j, k, \dots$  range from 0 to 2. We denote the 3-dimensional Levi-Civita tensor by  $\epsilon^{\mu\nu\sigma}$ and the Levi-Civita symbol as $\varepsilon^{012}=-1$ ($\epsilon_{\mu\nu\sigma}=\sqrt{-g}\varepsilon_{\mu\nu\sigma}$, $\epsilon^{\mu\nu\sigma}=(\sqrt{-g})^{-1}\varepsilon^{\mu\nu\sigma}$).  We choose gamma matrices with tangent space indices as: 
$\gamma^0=i\sigma^2\,,\, \gamma^1 =\sigma^3\,,\,\gamma^2=\sigma^1$, where $\sigma$'s are the Pauli matrices. The charge conjugation matrix is $C=\gamma^0$. We have $\bar{\lambda}=\lambda^{\dagger} C$
and a Majorana spinor satisfies $\lambda^*=-i\lambda$.}.
One then derives algebraic and differential conditions on them using the supersymmetry transformations \eqref{eq4}-\eqref{eq3}. It turns out that $SO(4)$ spinor indices can be split as $A=(a, \Tilde{a})$ with $a=\{1,2\}$ and $\Tilde{a}=\{3,4\}$ and one can choose a basis in which
\begin{equation}
    F^{ab}=-f\epsilon^{ab} \, , \, F^{a \Tilde{a}}= F^{\Tilde{a}\Tilde{b}}=0 \, .
\end{equation}
Consequently, only $V_\mu^{ab}$'s are non-zero, and we define the following vectors using them:
\begin{align} \label{vectors}
V_\mu=V_\mu^{11}+V_\mu^{22} \,, \quad
K_\mu=V_\mu^{11}-V_\mu^{22} \,, \quad
L_\mu=2V_\mu^{12} \,,
\end{align}
which satisfy
\begin{align}
&V^\mu K_\mu = V^\mu L_\mu=K^\mu L_\mu=0 \notag \,, \quad 
V_{[\mu}K_{\nu]} = \epsilon_{\mu\nu\sigma}f L^\sigma \,, \\
&V^\mu V_\mu=-K^\mu K_\mu=-L^\mu L_\mu=-4f^2 \,.
\label{algebraic}
\end{align}
It is easy to see that when $f \neq 0$, these three vectors form a 3-dimensional orthogonal basis \cite{Deger:2010rb}. Now, using the Fierz identity, supersymmetry breaking conditions are found as
\begin{align}
    V^\mu\gamma_\mu\lambda_a =2f \epsilon_{ab}\lambda^b \, , \quad K^\mu\gamma_\mu\lambda_a =2f \, (-1)^a \epsilon_{ab}\lambda^b \, , \quad L^\mu\gamma_\mu\lambda_a =2f\, (-1)^a \lambda^a \,  .
\label{break}
\end{align}
Only two of these relations are independent and there is no summation on the index $a$. As for the differential conditions, from \eqref{eq4}, one derives
\begin{align} \label{diff}
&\nabla_\mu V_\nu =-W\epsilon_{\mu\nu\sigma} V^\sigma \, , \notag\\
&\nabla_\mu K_\nu =-W\epsilon_{\mu\nu\sigma} K^\sigma +2X_\mu L_\nu \, , \notag\\
&\nabla_\mu L_\nu =-W\epsilon_{\mu\nu\sigma} L^\sigma -2X_\mu K_\nu \, .
\end{align}
The first one implies that $V$ is either a timelike ($f\neq 0$) or a null ($f=0$) Killing vector. Moreover, one gets
\begin{equation}
    \partial_\mu f=0 \, .
\end{equation}
Now, from the supersymmetry variations of matter sector \eqref{eq1}-\eqref{eq3} 
it can be shown that 
\begin{align} 
    \epsilon^{\nu\mu\sigma}\partial_\mu\xi_{1,2} \, V_\sigma+2f\,\epsilon^{\nu\sigma\rho}\, \mathcal{F}^{1,2}_{\rho\sigma} + 2\,V^\nu\,\frac{\partial W}{\partial \xi_{1,2}}    &= 0 \, , \label{YPV} \\
    f\, \partial_\nu\xi_{1,2} + \mathcal{F}^{1,2}_{\nu\sigma} \, V^\sigma &=0 \label{VF} \, , \\
     \epsilon^{\nu\mu\sigma}\partial_\mu\rho V_\sigma-2fg^{\mu\nu}\sinh\rho\,D_\mu\theta + 2  \,V^\nu \, \frac{\partial W}{\partial \rho} 
     &= 0 \, ,  \label{YPVrho} \\
2f\partial^\nu\rho+\sinh\rho\,\epsilon^{\nu\mu\sigma}\, D_\mu\theta \, V_\sigma &= 0 \, .  \label{epsiYrho}
\end{align}
Equations \eqref{eq1}-\eqref{eq3} additionally give 
\begin{align}
\mathcal{L}_V \xi_1=\mathcal{L}_V \xi_2=\mathcal{L}_V \rho=0  \, ,\label{scalarslie}
\end{align}
where $\mathcal{L}_V$ is the Lie derivative in the Killing direction $V$. Moreover, we have
\begin{align}
    \mathcal{L}_V \mathcal{A}_\mu^{1,2} = 0 \, ,
\label{LieA}
\end{align}
 after choosing the gauge 
\begin{align} 
V^\mu \mathcal{A}_\mu^{1,2}= -f\xi_{1,2} \label{gauge} \, .
\end{align}
For $\rho=0$, the scalar field $\theta$ completely drops out from the model \eqref{Lagrangian} and when $\rho \neq 0$, with this gauge choice one finds
\begin{align}
    \mathcal{L}_V \theta =2fg_0(2e^{\xi_1}+\xi_1) \, .
\label{thetaLie}
\end{align}

After this analysis, we are now ready to look for supersymmetric solutions of the model. Since the null case ($f=0$) was exhausted in \cite{Deger:2024xnd} for the gauged theory, we will focus on the timelike case (i.e. $f \neq 0$) in this paper. The null supersymmetric solutions of the ungauged model will be derived in the appendix \ref{a1}, which were not worked out in \cite{Deger:2024xnd}.

\section{Timelike Killing Vector} \label{4}
The most general 3-dimensional spacetime metric that admits $V=\partial_t$  as a Killing vector with the constant negative norm $-4f^2$ can be written as
\cite{Deger:2013yla}
\begin{align}
ds^2=-4f^2\bigl[dt+M(x,y)dx+N(x,y)dy\bigr]^2+e^{2\sigma(x,y)}(dx^2+dy^2) \, . \label{timelikemetric}
\end{align}
Note that, \, $\sqrt{-g} =2 \,f \, e^{2\sigma}$ where we assume $f>0$ without loss of generality.  Choosing vielbeins of the metric \eqref{timelikemetric} as 
\begin{align}
e^0=-2f(dt+Mdx+Ndy) \, , \quad  e^1=e^\sigma dx \, , \quad e^2=-e^\sigma dy \, , \label{timelikevielbeins}
\end{align}
non-zero spin connections are found to be
\begin{align}
\omega_t^{\;12}&=2e^{-2\sigma}f^2(M_y-N_x) \, , \qquad
\omega_x^{\;12} =-\partial_y\sigma+2e^{-2\sigma}f^2M(M_y-N_x) \, , \notag\\
\omega_x^{\;02}&=\omega_y^{\;01}=e^{-\sigma}f(M_y-N_x) \, ,\quad 
\omega_y^{\;12}=\partial_x\sigma+2e^{-2\sigma}f^2N(M_y-N_x) \, .
\end{align}

The algebraic conditions \eqref{algebraic} can be satisfied with
\begin{align}
K_\mu=(0,2fe^\sigma\sin\varphi,2fe^\sigma\cos\varphi)
 \, , \quad L_\mu=(0,2fe^\sigma\cos\varphi,-2fe^\sigma\sin\varphi) \, ,
\end{align}
where $\varphi(t,x,y)$ is an arbitrary function. We now solve the supersymmetry breaking conditions \eqref{break} with these $(V,K,L)$ vectors and get
\begin{align} \label{Killing}
\lambda_1=(1+i)\, Z \, \begin{bmatrix} 1-\cos\varphi \\ -\sin\varphi \end{bmatrix} \, , \quad \lambda_2= \gamma_0 \, \lambda_1 \, ,
\end{align}
where $Z(t,x,y)$ is a real valued function. It can be fixed from
\begin{align}
f=F^{12}=\bar\lambda^1\lambda^2=4(1-\cos\varphi)Z^2  \, ,
\end{align}
after which the Killing spinor is completely determined in terms of the function $\varphi(t,x,y)$ as
\begin{align} \label{Killing3}
   \lambda_1=(1+i)\, \sqrt{\frac{f}{2}} \begin{bmatrix} \sin(\varphi/2) \\ -\cos(\varphi/2) \end{bmatrix} \, , \quad \lambda_2=(1+i)\, \sqrt{\frac{f}{2}} \begin{bmatrix} \cos(\varphi/2) \\ \sin(\varphi/2) \end{bmatrix} \, .
\end{align}
Note that our Killing spinors are always smooth.

Meanwhile, the differential relations between these vectors \eqref{diff} imply that components of $X_\mu$ defined in \eqref{X} are fixed as  
\begin{align}\label{X2}
X_t&= 2Wf +\frac{1}{2} \partial_t\varphi \, , \nonumber\\
X_x&= 2WfM+\frac{1}{2}(\partial_y\sigma+\partial_x\varphi) \, , \nonumber\\
X_y&= 2WfN+\frac{1}{2}(-\partial_x\sigma+\partial_y\varphi) \, , 
\end{align}
and the superpotential $W$ \eqref{superpotential} satisfies
\begin{align}
W&=-fe^{-2\sigma}(M_y-N_x) \label{Weqn} \, . 
\end{align}
With these equations at hand, one can show that the Killing spinor equation \eqref{eq4} is satisfied identically for the Killing spinors we found in \eqref{Killing3}. 

Now, the BPS conditions \eqref{YPV} and \eqref{VF} give
\begin{align} \label{F}
&\mathcal{F}_{tx}^{1,2}=f\partial_x\xi_{1,2} \notag \, ,\\
&\mathcal{F}_{ty}^{1,2}=f\partial_y\xi_{1,2} \notag \, , \\
&\mathcal{F}_{xy}^{1,2}=-\frac{\partial W}{\partial \xi_{1,2}}e^{2\sigma}+f(\partial_y\xi_{1,2} M-\partial_x\xi_{1,2} N) \, . 
\end{align}
It is important to notice that the first two equations above are compatible with our gauge choice \eqref{gauge}. Next, from \eqref{YPVrho} and \eqref{epsiYrho} we get
\begin{align} \label{theta}
0&=\sinh\rho\; (D_t\theta-4fg_0e^{\xi_1}) \, , \notag\\
\partial_y \rho&= \sinh\rho\; (D_x\theta -4fMg_0e^{\xi_1} ) \, , \notag\\
 \partial_x \rho&= -\sinh\rho\; (D_y\theta -4f Ng_0e^{\xi_1} ) \, , 
\end{align}
where the first equation is nothing but \eqref{thetaLie}.
We now have all the information coming from BPS conditions \eqref{eq4}-\eqref{eq3} of our model for the timelike case. We can simplify things by combining \eqref{X2} with \eqref{F} and \eqref{theta} in the definition \eqref{X} of $X_\mu$ which results in
\begin{align} \label{phi}
\partial_t\varphi&=-2f\mathcal{H}\notag \, ,\\
\partial_y\big(\sigma-4(\xi_1+\xi_2)+\frac{1}{2}\ln(\cosh\rho+1)\big)+\partial_x\varphi&=-2fM\mathcal{H}\notag \, ,\\
\partial_x\big(\sigma-4(\xi_1+\xi_2)+\frac{1}{2}\ln(\cosh\rho+1)\big)-\partial_y\varphi&=2fN\mathcal{H} \, ,
\end{align}
where 
\begin{align} \label{H}
\mathcal{H}
&=9k_0e^{\xi_1+\xi_2}-9g_0e^{\xi_1}\cosh \rho-g_0e^{\xi_1} -10g_0e^{\xi_2} \, .
\end{align} 
We can now integrate the first equation of \eqref{phi} to find
\begin{align}
    \varphi = -2f\mathcal{H} t + G(x,y) \, ,
\end{align}
where $G(x,y)$ is an arbitrary function. Using this in the last two equations of \eqref{phi} immediately shows that $\mathcal{H}$ has to be a constant since only the function $\varphi$ can depend on time in them. When $\mathcal{H}\neq 0$ (which is the case here, as we will see below)
we can actually absorb $G(x,y)$ into the time coordinate by defining  $\hat t = t - G/(2f\mathcal{H})$. This changes $M\rightarrow \hat M=M- \partial_x\varphi/(2f\mathcal{H})$ and $N\rightarrow \hat N=N- \partial_y \varphi/(2f\mathcal{H})$ in the metric \eqref{timelikemetric} and $\partial_x\varphi$ and $\partial_y\varphi$ drops out from equations. Therefore,  we can set $G(x,y)=0$ or equivalently $\partial_x\varphi =\partial_y\varphi=0$ without loss of generality for $\mathcal{H}\neq 0$. After this crucial observation, we are now ready to start analyzing the field equations \eqref{Eineq}-\eqref{A2eqn} together with the supersymmetry conditions \eqref{Weqn}-\eqref{phi}.

\section{Supersymmetric Solutions of the Gauged Model} \label{5}
In this section, we consider the gauged model \eqref{Lagrangian}, for which constants $g_0$ and $k_0$ of the theory are assumed to be non-vanishing. Using BPS conditions \eqref{F} and \eqref{theta} in the vector field equations \eqref{A2eqn} we get
\begin{align}
k_0e^{\xi_1-\xi_2}-2g_0e^{-\xi_2}-k_0\xi_1&=c_1\notag\\
k_0e^{-\xi_1+\xi_2}-2g_0e^{-\xi_1}\cosh\rho-4g_0\cosh\rho-k_0\xi_2&=c_2 
\end{align}
for some $c_1,c_2\in\mathbb{R}$. Considering these together with the fact that $\mathcal{H}$ in \eqref{H} is also constant, it is easy to check that 
scalar fields $(\xi_1, \xi_2, \rho)$ have to be constant. But, then, from the remaining field equations one can show that the only possible choice for these constants is their value at the supersymmetric AdS vacuum \eqref{AdSvacuum}, which is perhaps not a surprise since that is the only supersymmetric critical point of the theory at which we have $\mathcal{H}= 2W=-4g_0^2/k_0$.
However, this choice sets all the field strengths to zero from \eqref{F} since $\frac{\partial W}{\partial \xi_{1,2}}=0$ at this point
and after these, \eqref{Eineq}  reduces to $R_{\mu\nu}= -\frac{8g_0^4}{k_0^2}g_{\mu\nu}$ which gives
\begin{align} \label{mn}
&\partial_y\sigma=\frac{8g_0^2}{k_0}fM  \, , \quad \partial_x\sigma=-\frac{8g_0^2}{k_0}fN  \, .
\end{align}
From these, using \eqref{Weqn} we find
\begin{align}
    \Box_{(x,y)}\sigma=\frac{16g_0^4}{k_0^2}e^{2\sigma} \, ,
\end{align}
which is Liouville's equation where $-16g_0^4/k_0^2$ is the Gaussian curvature of the hypersurface parametrized by the $(x,y)$. Its solution is
\begin{align}
\sigma=-\ln\bigg|1-\frac{4g_0^4}{k_0^2}(x^2+y^2)\bigg|  \, ,
\end{align}
from which $M$ and $N$ can be computed using \eqref{mn}. After that
the metric \eqref{timelikemetric} becomes
\begin{align}
ds^2=-\bigg(2fdt+\frac{2g_0^2}{k_0}\frac{(ydx-xdy)}{1-\frac{4g_0^4}{k_0^2}(x^2+y^2)}\bigg)^2+\frac{1}{\big[1-\frac{4g_0^4}{k_0^2}(x^2+y^2)\big]^2}(dx^2+dy^2) \, .
\end{align}
Making the change of coordinates
\begin{align}
\sqrt{x^2 + y^2} = \frac{k_0}{2g_0^2} \sin\alpha 
\, , \quad \tau=\frac{4 f g_0^2}{k_0}t \, \quad \gamma= \frac{4 f g_0^2}{k_0}t + \arctan\frac{y}{x} \, , 
\end{align}
in the above, we get
\begin{align}
ds^2=\frac{k_0^2}{4g_0^4 \cos^2\alpha}[-d\tau^2+d\alpha^2+\sin^2\alpha d\gamma^2] \, ,
\end{align}
which is the conformal metric for AdS$_3$ spacetime in global coordinates. Hence, we conclude that AdS$_3$  is the only supersymmetric solution for the gauged theory in the timelike class. It already appeared as a solution in the null Killing vector case \cite{Deger:2024xnd} but in the Poincar\'e patch.

\section{Supersymmetric Solutions of the Ungauged Model}\label{6}
We now consider the ungauged limit of our model \eqref{Lagrangian}
for which we must set $g_0=k_0=0$ which implies $W=\mathcal{H}=0$. Consequently, we can set $M=N=0$ in our spacetime metric \eqref{timelikemetric} since nothing depends on time, and these functions can be eliminated in the metric by making a shift in the $t$-coordinate \cite{Deger:2010rb}. So, for this case, our spacetime metric takes the form
\begin{align}
ds^2=-dt^2+e^{2\sigma(x,y)}(dx^2+dy^2) \, , \label{timelikemetric2}
\end{align}
where we fixed $f=1/2$ by rescaling the time coordinate.
Then, equations \eqref{F} reduce to
\begin{align} \label{F2}
\mathcal{F}_{tx}^{1,2}=f\partial_x\xi_{1,2}  \, ,\quad
\mathcal{F}_{ty}^{1,2}=f\partial_y\xi_{1,2}  \, , \quad
\mathcal{F}_{xy}^{1,2}= 0 \, . 
\end{align}
Meanwhile, equations \eqref{theta} become
\begin{align} \label{theta2}
\sinh\rho\; \partial_t\theta = 0 \, , \quad
\partial_y \rho= \sinh\rho\; \partial_x\theta  \, , \quad
 \partial_x \rho= -\sinh\rho\; \partial_y\theta  \, . 
\end{align}
Finally, equations \eqref{phi} simplify to
\begin{align} \label{phi2}
\partial_t\varphi&=0\notag \, , \\
\partial_y\big(\sigma-4(\xi_1+\xi_2)+\frac{1}{2}\ln(\cosh\rho+1)\big)+\partial_x\varphi&=0\notag \, ,\\
\partial_x\big(\sigma-4(\xi_1+\xi_2)+\frac{1}{2}\ln(\cosh\rho+1)\big)-\partial_y\varphi&=0 \, .
\end{align}
Unlike the gauged case, we can not make $\partial_x \varphi$ and $\partial_y \varphi$ vanish by a shift in the time coordinate since $\varphi$ does not depend on time. Now, field equations of $\xi_1$ and $\xi_2$ \eqref{xi1eqn}-\eqref{xi2eqn} give
\begin{align}
e^{-2\xi_{1,2}}\bigl[(\partial_x\xi_{1,2})^2+(\partial_y\xi_{1,2})^2\bigr]=4(\partial_x^2\xi_{1,2}+\partial_y^2\xi_{1,2}) \, .
\end{align}
Whereas, from the vector field equations \eqref{A1eqn}-\eqref{A2eqn} we find
\begin{align}
2(\partial_x\xi_{1,2})^2+2(\partial_y\xi_{1,2})^2=\partial_x^2\xi_{1,2} +\partial_y^2\xi_{1,2} \, ,
\end{align}
which is clearly in conflict with the preceding result unless we have
\begin{align}
    \xi_{1}=\xi_{2}=\textrm{constant} \implies \mathcal{F}_{\mu\nu}^{1,2} =0 \, ,
\end{align}
where the implication follows from \eqref{F2}. Now, one can check that all the remaining field equations are satisfied automatically once the BPS conditions above are fulfilled. After these, only equations \eqref{theta2} and \eqref{phi2} remain to be solved, which can be written in a more compact way as follows:
\begin{align}\label{compact}
&\partial_t\theta=\partial_t\varphi = 0  \, ,\\
&\partial_y\big(\ln(\tanh \frac{\rho}{2})\big)= \partial_x\theta 
\, , \quad \partial_x\big(\ln(\tanh \frac{\rho}{2})\big)=-\partial_y\theta \, , \label{bps1} \\
&\partial_y w =-\partial_x\varphi \, , \quad 
\partial_x w=\partial_y\varphi \label{bps2} \, ,
\intertext{where we defined}
&\sigma = w-\ln(\cosh \frac{\rho}{2})  \label{sigma} \, .
\intertext{Observe that these first order equations yield 4 harmonic functions}
&\Box\theta= 
\Box(\ln(\tanh \frac{\rho}{2})\big)=
\Box\varphi=\Box w=0 \label{harmonic} \, .
\end{align}
Let us first show that the metric function $\sigma(x,y)$ given in \eqref{sigma} is actually related to the K\"ahler potential of the sigma model target space. To see this explicitly, we will use the Poincar\'e disk model description of the hyperbolic plane without loss of generality. We now map the target space metric to the Poincar\'e disk metric  
\begin{align}\label{target}
    ds^2_{\textrm{target}}=\frac{1}{4} (d\rho^2+\sinh^2\rho\; d\theta^2) \, \longrightarrow \, ds^2_{\textrm{disk}}=\frac{d\zeta d\bar{\zeta}}{(1-\zeta \bar{\zeta})^2} \, ,
\end{align}
with $0 \leq \zeta\bar{\zeta}<1$ where
\begin{align}\label{zeta}
&\zeta=  \zeta_1 + i\zeta_2= \tanh \frac{\rho}{2} e^{-i\theta}  \, .
\end{align}
The K\"ahler potential $K(\zeta_1,\zeta_2)$ of the disk metric can now be found from $g_{\zeta \bar{\zeta}}=\partial_{\zeta}\partial_{\bar{\zeta}}K$ as
\begin{align}\label{Kahler}
    K=-\ln(1-\zeta\bar{\zeta}) =2\ln\cosh \frac{\rho}{2}
    \implies \sigma=w-\frac{K}{2}
    \, ,
\end{align}
which proves our assertion. In general, K\"ahler potential
is defined up to an additive harmonic function, and here, $K$ represents its non-harmonic part. Notice that in the metric function $\sigma(x,y)$ \eqref{sigma} we already have an arbitrary harmonic function $w$ \eqref{harmonic}. From \eqref{Kahler} we find 
\begin{align} \label{conformal}
    ds^2_{\textrm{disk}}=e^{2K} (d\zeta_1^2 + d\zeta_2^2) \, .
\end{align}
Moreover, $K$ satisfies Liouville's equation
\begin{align} \label{Liouville}
    \Box_{(\zeta_1,\zeta_2)} K= 4 e^{2K} \, ,
\end{align}
which shows that the Gaussian curvature of our Poincar\'e disk is -4 and not -1 which is due to the factor 1/4 that we put in the target space and the disk metric \eqref{target} for convenience. 

Now, using the complex coordinate $z=x+iy$, equations \eqref{bps1} and \eqref{bps2} can be solved in terms of arbitrary holomorphic functions $H_1(z)$ and $H_2(z)$ as follows: 
\begin{align}\label{h1}
    H_1(z) &= \ln(e^{-i \theta} \tanh \frac{\rho}{2})\equiv T(x,y) + iS(x,y)    \, , \\
    H_2(z) &= w(x,y) + i \varphi(x,y)  \, . \label{h2}
\end{align}
So, we can express our most general supersymmetric solution as
\begin{align}
    ds^2_{\textrm{spacetime} } &= -dt^2 + e^{2w} (1-e^{2T(x,y)}) (dx^2+ dy^2) \label{solnmetric}   \, ,\\
    \rho& = 2\, \textrm{arctanh}\, e^{T(x,y)}  \quad ,\quad   \theta= -S(x,y) \, , \label{solnscalar}
\end{align}
and the Killing spinors are given in \eqref{Killing3}. Here we used the identity: $e^{-K}=1-e^{2T}$. 

Actually, the holomorphic function $H_1(z)$ can be viewed as a map between the target space
$\mathbb{H}^2$ with the metric \eqref{target} to the $(x,y)$ coordinates with the Euclidean metric. To make this concrete, we first compute
\begin{align} \label{pde}
    \Box_{(x,y)} K = e^{2K} \Box_{(x,y)} e^{2T(x,y)}= e^{2K} \, 4  e^{2T(x,y)}[(\partial_x T)^2+(\partial_y T)^2] \, .
\end{align}
Since $T(x,y)$ is a harmonic function, the logarithm of the factor next to $e^{2K}$ above is a harmonic function too, since 
\begin{align}
\Box_{(x,y)} \ln [(\partial_x T)^2+(\partial_y T)^2] =0 \, .
\end{align} 
Therefore, \eqref{pde} can be rewritten as 
\begin{align}
    \Box_{(x,y)} \tilde{K} = 4 e^{2\tilde{K}} \, ,
\end{align}
where $\tilde{K}=K + \frac{1}{2}  \ln [\, \Box_{(x,y)} e^{2T(x,y)}] -\ln 2$,  so that the difference between K\"ahler potentials $K$ \eqref{Kahler}
and $\tilde{K}$ of $\mathbb{H}^2$ is a harmonic function. This equation is nothing but the standard Liouville equation as \eqref{Liouville}. Hence, similar to the disk mapping \eqref{target}, $(x,y)$ are conformally flat (or isothermal) coordinates for $\mathbb{H}^2$:
    \begin{align}\label{target2}
    ds^2_{\textrm{target}}=\frac{1}{4} (d\rho^2+\sinh^2\rho\; d\theta^2) \, \longrightarrow \, ds^2_{\textrm{space}}= e^{2\tilde{K}} (dx^2+dy^2) \, .
\end{align}
When $H_1(z)$ is not constant using this map in the opposite direction, we find
\begin{align}
    ds^2_{\textrm{spacetime}}= -dt^2 + \left(\frac{e^{2w}}{\Box_{(x,y)} e^{2T(x,y)} }\right) \, \frac{1}{\cosh^6\frac{\rho}{2}}\,  (d\rho^2+\sinh^2\rho\; d\theta^2) \, , \label{timelikemetric4}
\end{align}
which can also  be derived directly using \eqref{solnscalar} in \eqref{solnmetric}. It is remarkable that in this way the metric of the hyperbolic target space of the sigma model \eqref{target} became part of the spacetime metric \eqref{timelikemetric4} for any non-constant choice of $H_1(z)$. 
Hence, the warp factor in the spacetime metric \eqref{timelikemetric4}
can be  written as in \eqref{metricgen}, namely as $e^He^{-3K}$, where $H= 2w - \ln [\, \Box_{(x,y)} e^{2T(x,y)}] $ is a harmonic function.  

The curvature scalar of the spacetime metric \eqref{solnmetric} or \eqref{timelikemetric4} is
\begin{align} \label{curvaturescalar}
    R=e^{-2w+3K} \, \Box_{(x,y)} e^{2T(x,y)} \, ,
\end{align}
which is simply the inverse of the warp factor in \eqref{timelikemetric4}.

To summarize, our most general solution \eqref{solnmetric}-\eqref{solnscalar} is characterized in terms of two arbitrary holomorphic functions $H_1(z)$ and $H_2(z)$ defined in \eqref{h1}-\eqref{h2}. The only effect of the scalar fields on the spacetime metric \eqref{solnmetric} is via the K\"ahler potential $K$ of the sigma model target space which is determined in terms of spacetime coordinates by $H_1(z)$. These properties of our solutions match with those of the static supersymmetric solutions found for the $D=3, N=(2,0)$ supergravity coupled to a sigma model with an arbitrary K\"ahler target space in \cite{Howe:1995zm} (see its equations (4.24)-(4.26)) but we have an extra holomorphic function $H_2(z)$. Moreover, our sigma model target space is 2-dimensional which gives the possibility to identify it with the space part of the spacetime metric as we saw in \eqref{timelikemetric4}. We will now consider some special choices for these holomorphic functions for which the spacetime metric \eqref{solnmetric} and the Killing spinor \eqref{Killing3}  depend only on a radial coordinate.

\begin{itemize}
    \item \underline{$H_1(z)$=constant}
\end{itemize}

In this case, \eqref{h1} implies that scalar fields $\rho$ and 
$\theta$ are constant, which we take as zero without loss of generality. Thus, we are left with pure supergravity, and all solutions in this class are necessarily Ricci flat, i.e., $R_{\mu\nu}=0$  \eqref{Eineq}. An interesting case is when the metric \eqref{solnmetric} is only radial dependent which corresponds to choosing
\begin{align}
    H_2(z)= \ln (z) = k \ln(re^{i\beta}) \implies w= k \ln r \, , \quad \varphi=k\beta \, , 
\end{align}
where $r> 0$ and $0\leq |\beta| < 2\pi$ and we take $k\in [-1,0]$ to have an injective map. The metric of this solution \eqref{solnmetric} can be written as
\begin{align}
&ds^2=-dt^2+ du^2+u^2d\hat{\beta}^2 \, ,\quad (k\neq -1) \, , \\  
&ds^2=-dt^2+ dv^2+d\beta^2 \, ,    \quad (k=-1) \, . \label{Mink}
\end{align}
where $u=r^{k+1}/(k+1)$, $\hat{\beta}=|k+1|\beta$ and $v=\ln r$.
Obviously, $k=0$ corresponds to the Minkowski spacetime, but otherwise when  $k\neq -1$ this solution describes a conical spacetime
since the range of $\hat{\beta}$ is $0\leq |\hat{\beta}| \leq 2\pi |k+1| $. Finally, when $k=-1$, it is a cylinder with topology Minkowski$_2 \times \mathbb{S}^1$.

\begin{itemize}
    \item \underline{$H_1(z) = \kappa \ln (z)$}
\end{itemize}

Here, $\kappa \in \mathbb{R}\backslash \{0\}$.  Writing $z=x+iy=re^{i\beta}$  from 
\eqref{h1} we get 
\begin{align} \label{trans}
e^T=\tanh\frac{\rho}{2}= r^\kappa \, , \quad \theta= - \kappa \beta \, .
\end{align}
 For this to be well-defined, we need to restrict the radial coordinate as $0\leq r^\kappa < 1$ so that $0 \leq r < 1$ \, for \, $\kappa >0$ \, and \, $1< r < \infty$ \, for \, $\kappa<0.$ To make the identification between angular coordinates one-to-one we take their ranges as $0\leq \beta < 2\pi$, \, $0\leq |\theta| < 2\pi/|\kappa|$ \, for \, $0<|\kappa|\leq 1$ \, and \, $0\leq \beta < 2\pi/|\kappa|$, \, $0\leq |\theta| < 2\pi$ \, for \, $|\kappa|\geq 1$. Let us note that this case is similar to the disk model mapping \eqref{target}  since complex coordinate $\zeta$ of the disk model \eqref{zeta} and the spacetime complex coordinate $z$ are related via $\zeta=z^{\kappa}$. 
Now, with these  \eqref{solnmetric} becomes
\begin{align} \label{radial} 
ds^2&=-dt^2+e^{2w} (1-r^{2\kappa}) (dr^2+ r^2 d\beta^2) \, ,
\end{align}
which using transformations \eqref{trans} can also be written
in the form of \eqref{timelikemetric4} 
\begin{align}\label{hypermet}
   ds^2 &= -dt^2+e^{2w} \frac{(\tanh\frac{\rho}{2})^{\frac{2}{\kappa}-2}}{4 \kappa^2 \cosh^6\frac{\rho}{2}} (d\rho^2+\sinh^2\rho\; d\theta^2) 
\, .
\end{align}
  Also, note that the solution \eqref{radial} is only radial dependent when the function $w$ is also chosen like that. 
The scalar curvature  of \eqref{hypermet} can be read from \eqref{curvaturescalar} as
\begin{align}\label{curvature}
    R= 4 \kappa^2 e^{-2w} (\tanh \frac{\rho}{2})^{2-\frac{2}{\kappa}} \,\cosh^6 \frac{\rho}{2} \,  .
\end{align}
For $H_2(z)=$constant (which means that $w$ and the Killing spinor \eqref{Killing3} are constant) curvature
blows up as $\rho \rightarrow \infty$ (i.e., $r \rightarrow 1$). Note that this is a naked ring shaped singularity. There may also be a singularity as $\rho \rightarrow 0$ (i.e., $r^\kappa \rightarrow 0$) depending on the power of $\sinh(\rho/2)$. If $|\kappa|\geq 1$, so that the hyperbolic target space is completely covered, then $R$ approaches zero in this limit\footnote{In this limit for $\kappa=1$ the curvature scalar formula \eqref{curvature} has $0^0$ indeterminacy, and therefore it has to be computed separately first taking the limit in the metric.} and the spacetime geometry  
\eqref{radial} becomes
a cone if $\kappa>1$ and Minkowski for $\kappa=1$. Notice that these two limits of $\rho$ correspond to going to the boundary and the center of the Poincar\'e disk, respectively. Because the holomorphic function $H_2(z)$ \eqref{h2} remains free, there are infinitely many solutions. A natural choice is to take it in the same form with $H_1(z)$, that is $H_2=d_1\ln(z)$ where $d_1\in \mathbb{R}$ so that the solution remains only radial dependent and $\varphi$ becomes an angular coordinate in the spinor \eqref{Killing3}. Finally, let us point out that the metric \eqref{radial} with $\kappa=-1$ appeared as a solution to $D=3, N=8$ ungauged supergravity in \cite{Deger:2015tra} (see its equation (155)).

\begin{itemize}
    \item \underline{$H_1(z) = \kappa_2 z$}
\end{itemize}

Here $\kappa_2 \in \mathbb{R}\backslash \{0\}$. With this $H_1(z)$ from \eqref{h1} we have
\begin{align}
    e^T=\tanh\frac{\rho}{2}= e^{\kappa_2 x}\, , \quad \theta= - \kappa_2 y \, .
\end{align}
Now our spacetime metric \eqref{solnmetric} becomes
\begin{align}
    ds^2=-dt^2+\frac{e^{2w}}{\kappa_2^2}\left(\frac{1}{\hat{r}^2}-1\right) (d\hat{r}^2+ \hat{r}^2d\theta^2) \, , 
\end{align}
where $\hat{r}= e^{\kappa_2 x}$ and $\hat{r} \in [0,1]$. So, the metric is again only radial dependent if $w$ is also so. 


We can express this metric in hyperbolic coordinates as in \eqref{timelikemetric4} which leads to
\begin{align} \label{hypermet2}
ds^2= -dt^2+e^{2w} \frac{(\tanh\frac{\rho}{2})^{-2}}{4 \kappa_2^2 \cosh^6\frac{\rho}{2}} (d\rho^2+\sinh^2\rho\; d\theta^2) 
\, .
\end{align}
The solution has the same form with 
\eqref{hypermet}.
The curvature scalar \eqref{curvaturescalar} is the inverse of the warp factor as in \eqref{curvature}. For constant $w$
the only singular limit is $\rho \rightarrow \infty$ (i.e. $ \hat{r}\rightarrow 1$), which is again a ring singularity.  As $\rho \rightarrow \ 0$ (i.e. $ \hat{r}\rightarrow 0$) it is asymptotically flat like \eqref{Mink}.

\section{Conclusions} \label{7}
In this paper, we studied supersymmetric solutions of the model \eqref{Lagrangian} that admit a timelike Killing vector constructed from Killing spinors and thereby completed the work done in \cite{Deger:2024xnd} where supersymmetric solutions with a null Killing vector were obtained. We found that when the supergravity is gauged, AdS$_3$ is the only allowed solution for this class. This is
in big contrast to the null case where several different types of solutions were found. This might be due to the fact that what we considered is a truncation  \cite{Deger:2019jtl} of the model that comes from 6-dimensions \cite{Deger:2014ofa}. It remains to be seen whether the full theory allows a richer set of timelike solutions like spacelike or timelike warped AdS as it happened in certain on-shell \cite{OColgain:2015jlg, OColgain:2015dbh, Karndumri:2015sia}  and off-shell \cite{Deger:2013yla, Deger:2016vrn}  $D=3, N=2$ supergravities. The same question can also be addressed for the $D=3, N=8$ gauged supergravity \cite{Nicolai:2001ac} where supersymmetric solutions are known only for the ungauged case \cite{Deger:2010rb, Deger:2015tra}. 

On the other hand, the ungauged limit of our model admits infinitely many such solutions expressed in terms of two arbitrary holomorphic functions. The correspondence that we found between the 2-dimensional sigma model target manifold and the space part of the spacetime geometry \eqref{timelikemetric4} is very intriguing.  A complete classification of timelike supersymmetric solutions was also achieved in \cite{Deger:2010rb, Deger:2015tra} for the $D=3$, $N=8$, ungauged supergravity and the role of the Liouville's equation was highlighted (see also \cite{Moutsopoulos:2016jza}). However, the target space of the sigma manifold had dimension bigger than 2, and consequently, such an identification was not available. It would be interesting to investigate this phenomenon for other 2-dimensional K\"ahler target spaces and in particular for $\mathbb{S}^2=SU(2)/U(1)$ in ungauged \cite{Howe:1995zm} and gauged supergravities \cite{Deger:1999st, Abou-Zeid:2001inc} in $D=3$. This strategy may also lead to new supersymmetric solutions in higher dimensional supergravities, which allow target space geometries other than K\"ahler manifolds, such as quaternionic manifolds.

The fact that we were able to solve the problem of finding supersymmetric solutions of our ungauged model completely, and holomorphic functions play an important role, hints at a possible connection with 2-dimensional integrable systems. For the timelike solutions, dimensional reduction along the Killing direction could give insights into understanding this relationship better. It would also be interesting to investigate Bogomol'nyi bound for our solutions as was done for the $CP^n$ sigma model coupled to $D=3, N=2$ supergravity \cite{Edelstein:1995md}.  Studying their solitonic spectrum similar to the work of \cite{Forste:1996aj, Forste:1997ac}
where $D=3, N=(2,0)$ ungauged supergravity coupled to sigma models $SL(2)/U(1)$ and $SU(2)/U(1)$ are considered, is desirable too. Finally, we would like to determine the higher dimensional origin of our ungauged model to be able to uplift the new solutions that we found. This was done, for example, 
for the static supersymmetric solutions of the $D=3, N=8$ supergravity that comes from the heterotic string theory compactified on 7-torus in \cite{Bakas:1997jy, Bourdeau:1997gm}.

\section*{Acknowledgments}
We thank Jan Rosseel and Henning Samtleben for useful discussions.
NSD and CAD are partially supported by 
the Scientific and Technological Research Council of T\"urkiye (T\"ubitak) project 123N953. NSD is grateful to ESI, Vienna, for hospitality during the final phase of this paper.

\appendix

\section{Null supersymmetric solutions of the ungauged model}
\label{a1}
Supersymmetric solutions of the gauged model \eqref{Lagrangian} with a null Killing vector were studied in \cite{Deger:2024xnd}, but those of the ungauged theory were left out, which we now derive by adapting its general analysis to $g_0=k_0=0$ with the same conventions. Our results in section \ref{3} are valid with $f=0$ and with the vector bilinears \eqref{vectors} chosen as $V^\mu=K^\mu$ and $L^\mu=0$.

The general spacetime metric that admits $V=\partial_v$ as a null Killing vector is \cite{Deger:2010rb, Deger:2024xnd}
\begin{align}
ds^2=dr^2+2dudv+e^{2\omega(u,r)}du^2 \, . \label{Gmetric}
\end{align}
For this metric, the only non-zero component of the Ricci tensor is
\begin{align}
    R_{uu}=-\frac{1}{2} \partial_r^2(e^{2\omega}) \, ,
\end{align}
which implies $R=0$. All physical fields are independent of the $v$-coordinate due to \eqref{scalarslie} and \eqref{thetaLie}. Our gauge choice \eqref{gauge} implies $\mathcal{A}_{v}^{1,2}=0$. A further gauge transformation can be made to set $\mathcal{A}_{r}^{1,2}=0$ so that vector fields have only $A_u^{1,2}$ component remaining. The Killing spinor equation \eqref{eq4} is solved using the supersymmetry breaking condition \eqref{break} as
\begin{align}
    \lambda^a= (1+i)\, e^{-\frac{\beta}{2}}\, \lambda_0^a \, ,
\end{align}
where $\lambda_0^a$ is a real, constant spinor that satisfies $(\gamma^1-\gamma^0)\lambda_0^a=0$. Additionally, one finds
\begin{align}\label{xcondition}
X_\mu=0  \implies
(1-\cosh\rho)\partial_u\theta+16(\mathcal{F}_{ru}^1+\mathcal{F}_{ru}^2)=0 \, ,
\end{align}
where the implication comes from $X_u=0$ \eqref{X}. Whereas $X_v=X_r=0$ are automatically satisfied. Now, from the BPS conditions \eqref{YPV}-\eqref{epsiYrho} we get
\begin{align}
\partial_r\xi_1=\partial_r\xi_2=\partial_r\rho=\sinh\rho\;\partial_r\theta=0 \, .  \label{scalarsrderivnungauged}
\end{align}
With these, scalar field equations \eqref{xi1eqn}-\eqref{thetaeqn} are satisfied identically. Next, from the vector field equations \eqref{A1eqn}-\eqref{A2eqn} and \eqref{xcondition} we find
\begin{align}
\mathcal{A}^1_u= Q(u)\, r \, , \quad
\mathcal{A}^2_u= -\left[ Q(u)+\frac{1}{16}(1-\cosh \rho)\partial_u\theta \right]r \, ,
\end{align}
where $Q(u)$ is arbitrary. Finally, solving Einstein's equation \eqref{Eineq}, the metric of the solution is found to be
\begin{align} \label{nullmetric}
ds^2=dr^2+ 2dudv+  \left( f_1(u)r+f_2(u) - \frac{P^2}{2} \right) du^2 \, ,
\end{align}
where $f_1(u)$ and $f_2(u)$ are arbitrary and
\begin{align}
    P^2= r^2 \left[ (\partial_u\xi_1)^2+(\partial_u\xi_2)^2 +(\partial_u\rho)^2+ \sinh^2\rho (\partial_u\theta)^2\right] +e^{-2\xi_1} (\mathcal{A}^1_u)^2 + e^{-2\xi_2} (\mathcal{A}^2_u)^2  \, .
\end{align}
The metric \eqref{nullmetric} describes a pp-wave in the Minkowski spacetime. Its form is exactly the same with the null supersymmetric solutions of the $D=3, N=8$ ungauged supergravity  \cite{Deger:2014ofa}. We verified that the $f_1(u)$ and $f_2(u)$ terms can actually be generated using Garfinkle-Vachaspati method \cite{Garfinkle:1990jq, Garfinkle:1992zj} (see \cite{Deger:2024xnd} for details) and therefore they are locally redundant \cite{Gibbons:2008vi, Chow:2009km}.

\bibliographystyle{utphys} 
\bibliography{references.bib}

\end{document}